\documentclass[prb,twocolumn,showpacs]{revtex4-1}
\usepackage[T1]{fontenc}
\usepackage{graphicx}
\usepackage{color}
\usepackage[dvipsnames]{xcolor}
\usepackage{dcolumn}
\usepackage{amsmath}
\usepackage{amssymb}
\usepackage{amsfonts,dsfont,xfrac}
\usepackage[bbgreekl]{mathbbol}
\usepackage{subfigure}
\usepackage{braket}
\usepackage{simplewick}
\usepackage{comment}

\usepackage{mathrsfs}
\usepackage{relsize,scalerel}
\usepackage{makecell}
\usepackage{placeins,float}

\usepackage[colorlinks=true, citecolor=black, linkcolor=black, urlcolor=black]{hyperref}
\usepackage{cleveref}

\newcommand{\eq}{\sss{\textup{eq}}}

\newcommand{\bR}{\boldsymbol{R}}

\newcommand{\bu}{\boldsymbol{u}}

\newcommand{\bk}{\boldsymbol{k}}

\newcommand{\sss}[1]{\scriptscriptstyle{{#1}}}

\newcommand{\Nat}{N_{\sss{a}}}
\newcommand{\Nc}{N_{{c}}}

\newcommand{\gn}[3]{\overset{\sss{(#1)}}{g}{}_{#2}^{#3}}

\newcommand{\Avg}[1]{\left\langle{#1}\right\rangle}
\newcommand{\AAvg}[1]{\left\langle\!\!\!\left\langle{#1}\right\rangle\!\!\!\right\rangle}

\newcommand{\avg}[1]{\langle{#1}\rangle}
\newcommand{\bReq}{\bR_{\eq}}

\newcommand{\ssst}[1]{\scriptscriptstyle{\text{#1}}}

\newcommand{\Vks}{V^{\ssst{KS}}}
\newcommand{\Vkshat}{\hat{V}{}^{\ssst{KS}}}
\newcommand{\whVks}{\hat{V}{}^{\ssst{KS}}}
\newcommand{\hVks}{\hat{V}{}^{\ssst{KS}}}

\newcommand{\Avgeq}[1]{\Avg{#1}_{\eq}}
\newcommand{\avgeq}[1]{\avg{#1}_{\eq}}

\newcommand{\gavg}{\bigl\langle\mkern-1mu g \mkern-1mu \bigr\rangle{}}
\newcommand{\gavgx}[1]{\bigl\langle\mkern-1mu \overset{\sss{(#1)}}{g} \mkern-1mu \bigr\rangle{}}
\newcommand{\gavgn}{\gavgx{n}}

\newcommand{\gx}[1]{\overset{\sss{(#1)}}{g}{}}
\renewcommand{\gn}{\gx{n}}

\newcommand{\alphasqF}{\alpha^2\!F}

\newcommand{\lambdan}{\overset{\sss{(n)}}{\lambda}}
\newcommand{\lambdax}[1]{\overset{\sss{(#1)}}{\lambda}}
\newcommand{\lambdauno}{\lambdax{1}}

\newcommand{\Nf}{N_{\sss{F}}}

\newcommand{\bhxi}[1]{\boldsymbol{\hat{\bi}}}

\begin{document}
\title{Enhanced superconductivity in palladium hydrides by non-perturbative electron-phonon effects}

\author{Raffaello Bianco$^{1,2,3,4}$}
\email{raffaello.bianco@unimore.it}
\author{\DJ{}or\dj{}e Dangi\'c$^{1,5}$}
\email{dorde.dangic@ehu.eus}
\author{Ion Errea$^{1,5,6}$}
\email{ion.errea@ehu.eus}

\affiliation{$^1$Centro de F\'isica de Materiales (CFM-MPC), CSIC-UPV/EHU,  Manuel de Lardizabal pasealekua 5, 20018 Donostia/San Sebasti\'an, Spain}

\affiliation{$^2$ Ru\dj er Bo\v{s}kovi\'c Institute, 10000 Zagreb, Croatia}

\affiliation{$^3$ Dipartimento di Scienze Fisiche, Informatiche e Matematiche, Universit\`a di Modena e Reggio Emilia, Via Campi 213/a I-41125 Modena, Italy}

\affiliation{$^4$ Centro S3, Istituto Nanoscienze-CNR, Via Campi 213/a, I-41125 Modena, Italy}

\affiliation{$^5$Fisika Aplikatua Saila, Gipuzkoako Ingeniaritza Eskola, University of the Basque Country (UPV/EHU), Europa Plaza 1, 20018 Donostia/San Sebasti\'an, Spain}

\affiliation{$^6$Donostia International Physics Center (DIPC), Manuel Lardizabal pasealekua 4, 20018 Donostia/San Sebasti\'an, Spain}

\begin{abstract}
Palladium hydrides exhibit the largest isotope-effect anomaly in superconductivity: replacing hydrogen with heavier isotopes increases the superconducting critical temperature. Although this behavior is commonly attributed to strong anharmonic hydrogen vibrations, \textit{ab initio} treatments have so far incorporated anharmonic effects only through phonon renormalization, neglecting non-linear contributions to the electron–phonon interaction vertices. While such approaches reproduce the anomalous isotope trend, they severely underestimate the critical temperatures.
Here, we show that non-linear electron–phonon coupling is essential in palladium hydrides. A straightforward inclusion of higher-order perturbative terms leads to a qualitative breakdown: the critical temperature is overestimated and the isotope anomaly is lost. We therefore adopt a non-perturbative framework based on an explicit evaluation of the ion-mediated electron–electron interaction, enabling anharmonic effects to be treated consistently in both the phonon spectra and the interaction vertices. Applied to PdH and PdD, it restores the anomalous isotope effect and brings calculated critical temperatures into a significantly improved agreement with experiments.  
\end{abstract}


\maketitle


Palladium hydride is one of the most fascinating and intriguing superconductors found up to date. 
It is not only one of the few hydrides that superconducts at ambient pressure~\cite{Flores-Livas2020APerspective,Pickard2020Superconducting,Bi2019TheSearch}, it displays the largest isotope anomaly known: the heavier the isotope of hydrogen, the larger the superconducting critical temperature~\cite{stritzker1972superconductivity,schirber1974concentration,SCHIRBER1984837}. 
This is in complete contradiction with the standard picture of electron-phonon mediated superconductors, where the critical temperature scales as $T_c\simeq\overline{\omega}\, e^{-1/\lambda}$, with $\overline{\omega}$ the characteristic phonon frequency and $\lambda$ the electron-phonon coupling constant. Considering that $\lambda\propto \Nf\,g^2/(M\overline{\omega}^2)$, where $\Nf$ is the density of states at the Fermi level, $g$ the mass-independent linear electron-phonon coupling vertex, $M$ the atomic mass, and that phonon frequencies in the standard harmonic approximation scale as $\overline{\omega}\propto M^{1/2}$, the electron-phonon coupling constant is expected to be the same for all isotopes. The mass dependence of the superconducting critical temperature is thus determined solely by $\overline{\omega}$, which decreases the heavier the isotope. Hence, the experimental fact that the $T_c$ of PdH, the hydrogen compound, is around $\sim$ 9 K~\cite{stritzker1972superconductivity,schirber1974concentration}, of PdD, the deuterium compound, $\sim$ 11 K, and apparently even higher for the tritium compound PdT~\cite{SCHIRBER1984837} is a big surprise.


There is broad consensus that the strong lattice anharmonicity of PdH lies at the origin of this anomaly~\cite{Ganguly1973,allen1983theory,ErreaB2013}. Indeed, density-functional theory (DFT) calculations including non-perturbative anharmonic effects within the stochastic self-consistent harmonic approximation (SSCHA)~\cite{Errea2014,Bianco2017,Monacelli2018A,Monacelli_2021} yield phonon spectra in very good agreement with inelastic neutron scattering and Raman experiments~\cite{Rowe1974Lattice,Rowe1986Isotope,Kolesnikov1991Neutron,Sherman1977Raman,Ross1998Strong}. These studies show that anharmonicity induces a substantial hardening of the hydrogen-derived optical modes throughout the Brillouin zone, particularly for vibrations associated with hydrogen atoms occupying the octahedral interstitial sites of the Pd face-centered cubic lattice. Owing to its lower mass, this anharmonic hardening is more pronounced in PdH than in PdD. Since the electron–phonon coupling constant $\lambda$ is inversely related to phonon frequencies, the stronger hardening in PdH suppresses $\lambda$ more effectively than in PdD, resulting in a larger coupling strength for the heavier isotope~\cite{ErreaB2013}. Including this anharmonic renormalization in the evaluation of $\lambda$ therefore naturally accounts for the anomalous isotope effect.

Combining the anharmonic SSCHA phonon spectra with the standard linear electron-phonon coupling vertices, the critical temperature of PdH, PdD, and PdT was estimated \emph{ab initio}. Even when using an unusually small Coulomb pseudopotential parameter to account for the electron–electron repulsion \cite{PhysRev.125.1263} ($\mu^*=0.085$), which tends to overestimate the critical temperature \cite{Flores-Livas2020APerspective}, significantly lower values than the experimental ones were predicted: 5.0 K for PdH, 6.5 K for PdD, and 6.9 K for PdT~\cite{ErreaB2013}. This discrepancy suggests that the different treatment of phonon frequencies—fully renormalized by anharmonicity—and electron–phonon interaction vertices—treated only at the lowest linear order in the ionic displacements—is inconsistent and leads to inaccurate predictions of the critical temperature.

In this Letter, by combining SSCHA anharmonic phonon spectra with a novel method to account for non-perturbative effects on the electron–phonon interaction~\cite{bianco2023nonperturbative}, presented in a companion manuscript submitted in parallel, we demonstrate that electrons couple nonlinearly to lattice vibrations in palladium hydrides. Higher-order derivatives of the electronic potential with respect to ionic displacements are found to be comparable to, and even larger than, the linear term, leading to a strong enhancement of the superconducting critical temperature.
Remarkably, when these nonlinear interaction vertices are treated as simple higher-order derivatives of the potential within a perturbative framework, the calculated critical temperature is largely overestimated and the isotope inversion in $T_c$ is lost. In contrast, when employing averaged electron–phonon vertices—obtained through the summation of an infinite class of Feynman diagrams and incorporating ionic fluctuations around the equilibrium positions—the resulting $T_c$ is brought into close agreement with experiment and the anomalous isotope effect is restored.
Our results demonstrate that a proper understanding of the critical temperature and the inverse isotope effect in palladium hydrides requires not only anharmonic renormalization of phonon frequencies, but also strong non-perturbative effects in the electron–phonon interaction itself.

\begin{figure}[t!]
\includegraphics[width=\columnwidth]{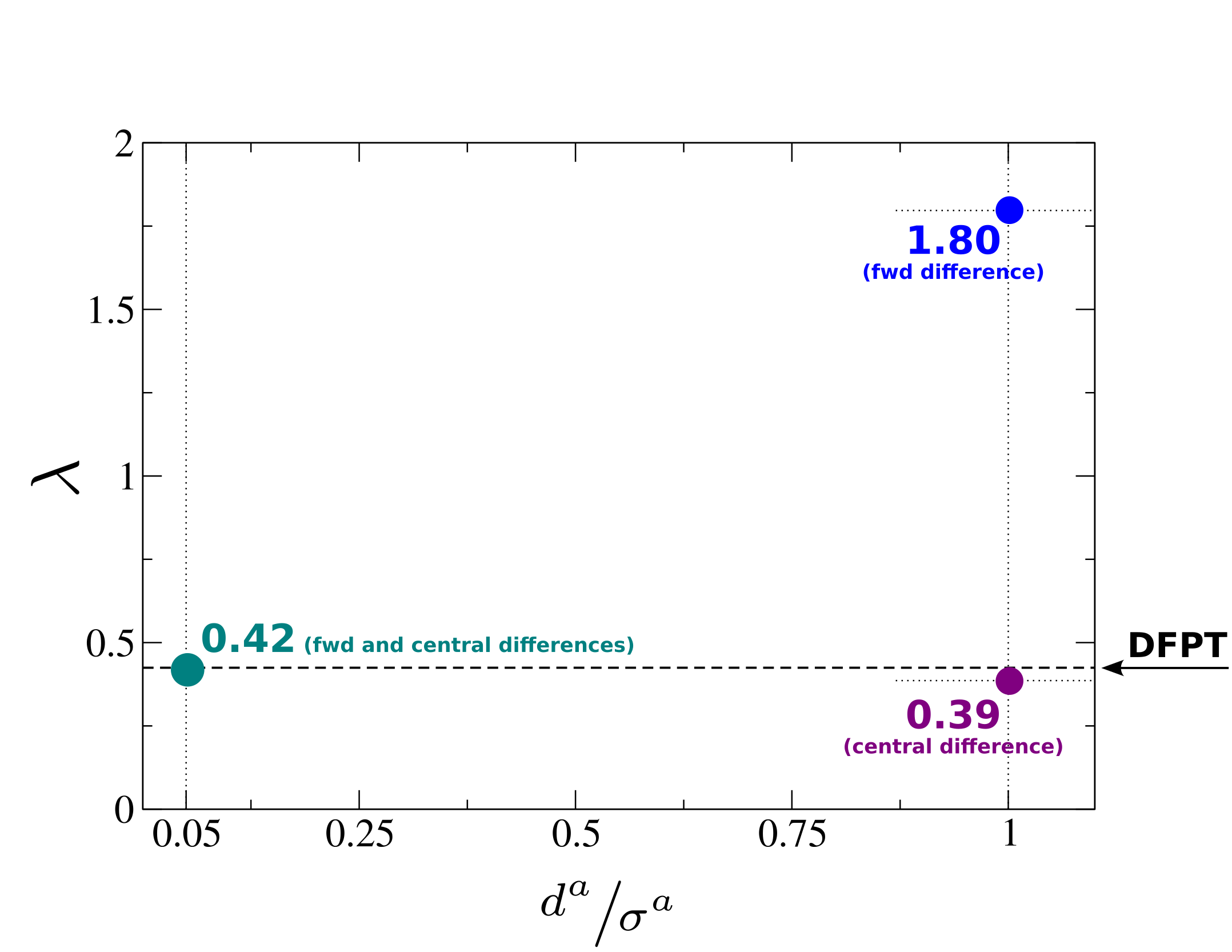}
\caption{ 
First-order electron-phonon coupling constant $\lambda$ of PdH calculated by finite differences, both with forward (Eq. \eqref{eq:fwd_derivative}) and central differences (Eq. \eqref{eq:central_derivative}), for different $d^a$ steps, making use of SSCHA phonon frequencies and polarization vectors.    
On the horizontal axis the ratio between $d^a$ and equilibrium root-mean-square displacement $\sigma^a=\sqrt{\avgeq{(u^a)^2}}$ along the same direction is indicated (the same ratio
$d^a/\sigma^a$ is considered for all the $a$). The horizontal dashed line represents the value of $\lambda$ obtained with DFPT.}
\label{fig:LambdaFD_PdH}
\end{figure}

\begin{figure*}[t!]
\includegraphics[width=\textwidth]{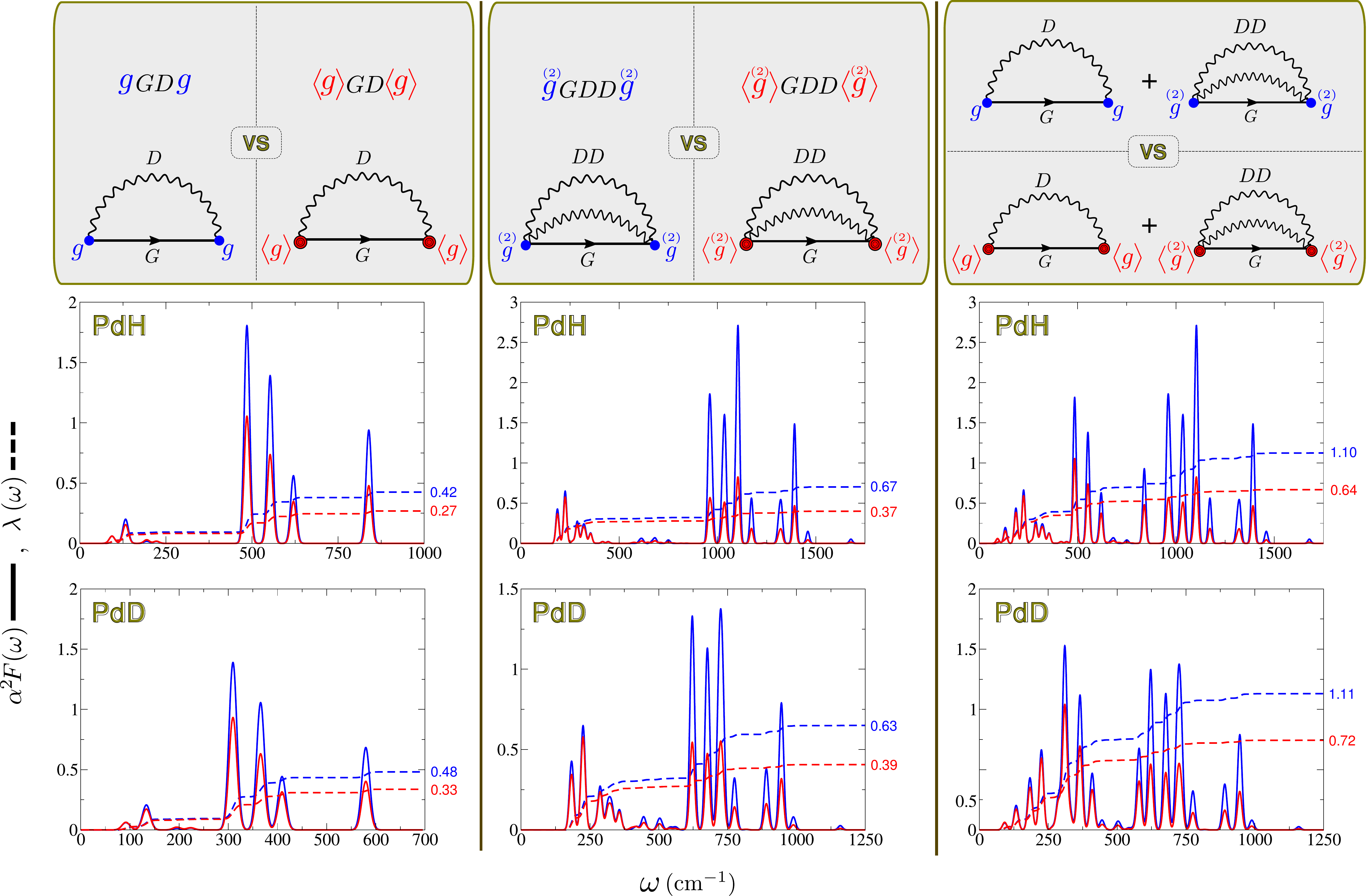}
\caption{(Left) Electron self-energy diagrams constructed with first-order bare $g$ vertices [Eq.~\eqref{eq:def_g_normal}] and averaged first-order $\gavg$ vertices [Eq.~\eqref{eq:def_gavgn_normal}]. $G$ and $D$ denote electron and phonon propagators, respectively. The corresponding self-energies are labeled $gGDg$ and $\gavg GD \gavg$. Below, the first-order $\overset{\sss{(1)}}{\alphasqF}(\omega)$ and $\lambdauno(\omega)$ calculated with $gGDg$ (blue) and $\gavg GD \gavg$ (red), for both PdH and PdD using SSCHA anharmonic phonon propagators. The final value of $\lambdauno$ is indicated. 
(Middle) Same analysis for the second-order contribution: $\overset{\sss{(2)}}{\alphasqF}(\omega)$ and $\lambdax{2}(\omega)$ computed using bare second-order vertices $\overset{\sss{(2)}}{g}$ and averaged second-order vertices $\gavgx{2}$. The corresponding self-energies are labeled $\overset{\sss{(2)}}{g}GDD\overset{\sss{(2)}}{g}$ and $\gavgx{2}GDD\gavgx{2}$. The resulting $\lambdax{2}$ is indicated. 
(Right) Total spectral functions and coupling strengths obtained from the sum of first- and second-order contributions. The final value $\lambda=\lambdauno+\lambdax{2}$ is reported.} 
\label{fig:main}
\end{figure*}



A DFT-based calculation of the electron-phonon interaction requires to compute the derivatives of the electronic $\hVks$ Kohn-Sham (KS) potential with respect to ionic displacements $u^a$ ($a$ labels both an atom and a Cartesian direction) from the equilibrium position $\bReq$. The standard isotropic linear-order Eliashberg function, calculated in this case in a supercell formed by $\Nc$ unit cells, is given by
\begin{equation}
\alphasqF(\omega)=
\Nc\Nf n! \sum_{\mu}
\AAvg{\,\left|g^{mn\bk}_{\mu}\right|^2\,}_{\substack{{m\bk}\\{n\bk}}\in\text{FS}}
\delta\left(\omega-\omega_{\mu}\right)
\label{eq:eli_linear}
\end{equation}
with
\begin{equation}
g^{mn\bk}_{\mu}=
\sum_a \frac{e_{\mu}^a}{\sqrt{2M_a\omega_{\mu}}}\left.\frac{\partial \Vks_{mn\bk}}{\partial u^a}\right|_{\bReq}.
\label{eq:linear_g}
\end{equation}
Here $\Vks_{mn\bk}=\braket{m\bk|\hVks(\bReq+\bu)|n\bk }$, with $\ket{m\bk}$ and $\ket{n\bk}$ the Bloch states of the undistorted system, and $\AAvg{\,\cdot\,}_{\substack{{m\bk}\\{n\bk}}\in\text{FS}}$ denotes the average over the Fermi surface (FS)~\cite{bianco2023nonperturbative} taken over the band indices $m$ and $n$ and the wave vector $\bk$ (the same for both states, as calculations are performed considering atomic distortions with supercell periodicity). In the equations above, $M_a$ is the mass of atom $a$, while $\omega_{\mu}$ and $e_{\mu}^a$  are the harmonic or SSCHA auxiliary phonon frequencies and polarization vectors of mode $\mu$, depending on whether anharmonic effects are neglected or treated non-perturbatively.
The integrated electron-phonon coupling is calculated as $\lambda(\omega)=\int_0^{\omega}d\omega'2\alphasqF(\omega')/\omega'$, whose high-frequency limit gives the electron-phonon coupling constant $\lambda=\int_0^{\infty}d\omega 2\alphasqF(\omega)/\omega$.
The linear-order derivative of $\hVks$ needed in the equations above can be calculated \emph{ab initio} routinely today both with density-functional perturbation theory (DFPT)~\cite{RevModPhys.73.515} or by taking finite-differences~\cite{frederiksen2007inelastic}. 

Finite difference calculations offer a simple way of estimating whether the electron-phonon coupling constant is linear or not. In order to calculate $\partial \whVks / \partial u^a|_{\bReq}$, the simplest approach is to displace the atoms from $\bReq$ as $\bReq  +  d^a\boldsymbol{\hat{a}}$, where $\boldsymbol{\hat{a}}$ is the $3\Nat$-dimensional unit vector along the direction $a$, so that 
\begin{equation}
    \partial \whVks / \partial u^a|_{\bReq} \sim [\Vkshat(\bReq +  d^a\boldsymbol{\hat{a}}) - \Vkshat(\bReq)]/d^a.
    \label{eq:fwd_derivative}
\end{equation}
We will refer to this derivative as the forward difference derivative, whose error is $\mathcal{O}(d^2)$. The calculation can be made more accurate, with error of $\mathcal{O}(d^3)$, by taking a central difference derivative:
\begin{equation}
    \partial \whVks / \partial u^a|_{\bReq} \sim [\Vkshat(\bReq +  d^a\boldsymbol{\hat{a}}) - \Vkshat(\bReq -  d^a\boldsymbol{\hat{a}})]/(2d^a).
    \label{eq:central_derivative}
\end{equation}
Both finite-difference estimates coincide with the exact derivative in the small-displacement limit. If the electron–phonon interaction is strictly linear, the central and forward finite-difference schemes yield the exact derivative for any value of $d^a$. In the presence of nonlinear contributions, however, the two estimates deviate as $d^a$ increases: the central scheme cancels higher-order even terms in the Taylor expansion of $\Vkshat$ with respect to the ionic displacements, whereas the forward scheme retains them.

Figure~\ref{fig:LambdaFD_PdH} shows that, although both approaches agree with the DFPT result for PdH at small displacements, they differ substantially within the physically relevant range of lattice distortions set by the root-mean-square ionic displacement. It is therefore internally inconsistent to adopt a linear electron–phonon vertex evaluated at equilibrium while the nuclei explore large anharmonic fluctuations. This inconsistency explains the underestimation of $\lambda$ reported in previous calculations, which combined anharmonic phonon spectra with a linear-order approximation of the electron–phonon coupling~\cite{ErreaB2013}. Moreover, the fact that the forward scheme enhances $\lambda$ whereas the central scheme suppresses it reveals a competition between perturbative orders with non-trivial sign dependence, rendering any finite-order truncation intrinsically uncontrolled and highlighting the need for a non-perturbative treatment of the electron–phonon interaction.

To address this issue, we have developed a novel non-perturbative \emph{ab initio} approach to compute the electron-phonon interaction, which includes non-linear effects and accounts for the quantum nature of the nuclei~\cite{bianco2023nonperturbative}. The method is based on the evaluation of the $GW^{\ssst{ph}}$ electron self-energy, where $W^{\ssst{ph}}$ is the effective nuclei-mediated electron-electron interaction, assuming that the nuclear dynamics are described by a Gaussian distribution. This approach defines a consistent prescription for the infinite resummation of non-linear electron-phonon contributions, rather than an explicit perturbative order counting, and the combination with the SSCHA, which is based on a Gaussian ionic distribution function, allows for a fully consistent non-perturbative treatment of both the phonon spectrum and the electron-phonon interaction. Within this framework, the isotropic Eliashberg function becomes an infinite series, $\alphasqF(\omega)=\sum_n\overset{\sss{(n)}}{\alphasqF}(\omega)$, where hte $n$th-term in the expansion is
\begin{align}
\overset{\sss{(n)}}{\alphasqF}(\omega)=
&\Nc\Nf\,n!\,\sum_{\mu_1\dots\mu_n}
\AAvg{\,\left|\gavgn^{mn\bk}_{\mu_1\ldots \mu_n}\right|^2\,}_{\substack{{m\bk}\\{m\bk}}\in\text{FS}}\nonumber\\
&\mkern100mu\times\delta\left(\omega-\sum_{h=1}^n\omega_{\mu_h}\right)\,,
\label{eq:eli_notimpl}
\end{align}
from which we define the $n$th-order contribution to the integrated electron-phonon coupling, $\lambdan(\omega)$, and to the electron-phonon coupling constant $\lambdan$.
At the diagrammatic level, the term~\eqref{eq:eli_notimpl} corresponds to the contribution to the electron self-energy arising from the $n$th-order electron–phonon Fan-like diagram, involving $n$ phonon propagators mediating the interaction between electrons. However, the $n$th-order electron–phonon vertex entering this contribution does not simply reduce to the $n$th derivative of the Kohn–Sham potential with respect to the ionic displacements,
\begin{align}
\gn^{mn\bk}_{\mu_1\ldots \mu_n}&=
\sum_{a_1\cdots a_n}
\frac{e^{a_1}_{\mu_1}}{\sqrt{2M_{a_1}\omega_{\mu_1}}}
\cdots 
\frac{e^{a_n}_{\mu_n}}{\sqrt{2M_{a_n}\omega_{\mu_n}}} \nonumber \\ & \times
\frac{\partial^n\Vks_{mn\bk}}{\partial u^{a_1}\ldots \partial u^{a_n}}\Big|_{\bReq}\,,
\label{eq:def_g_normal}
\end{align}
as in standard perturbation theory (see, e.g., Eqs.~\eqref{eq:eli_linear} and~\eqref{eq:linear_g} for the $n=1$ case), but rather by their quantum statistical averages,
\begin{align}
\gavgn^{mn\bk}_{\mu_1\ldots \mu_n}&=
\sum_{a_1\cdots a_n}
\frac{e^{a_1}_{\mu_1}}{\sqrt{2M_{a_1}\omega_{\mu_1}}}
\cdots 
\frac{e^{a_n}_{\mu_n}}{\sqrt{2M_{a_n}\omega_{\mu_n}}} \nonumber \\ & \times
\Avgeq{\frac{\partial^n\Vks_{mn\bk}}{\partial u^{a_1}\ldots \partial u^{a_n}}},
\label{eq:def_gavgn_normal}
\end{align}
where the average is taken with the Gaussian nuclear density matrix of the equilibrium ionic configuration (the SSCHA density matrix, in this case).
The averaged vertices $\gavgn$ incorporate the effect of ionic fluctuations on the electron–phonon interaction and exhibit a particularly transparent diagrammatic structure: they are obtained by dressing the corresponding bare interaction vertex $\gn$ with an infinite series of additional “flower” diagrams. Remarkably, this vertex renormalization systematically suppresses the electron–phonon coupling relative to its bare value, establishing lattice fluctuations as an intrinsic mechanism that weakens the electron–phonon interaction~\cite{bianco2023nonperturbative}.

\begin{table}[t!]
\centering
\begin{tabular}{l c c c} 
 \hline
 \hline
PdH, \ \ \ $T_c^{\mathrm{exp}} =$ 9 K\cite{stritzker1972superconductivity}, 8 K\cite{schirber1974concentration}  \\
\hline
Self-energy & $\lambda$ & $\omega_{\mathrm{log}}$ (cm$^{-1}$) & $T_c$ (K)  \\ 
\hline
$gGDg$ & 0.42 & 405.4 & 1.6  \\
$\gavg GD \gavg$ & 0.27 & 353.5 & 0.01 \\
$gGDg+\overset{\sss{(2)}}{g}GDD\overset{\sss{(2)}}{g}$& 1.10 & 497.0 & 49.6 \\
$\gavg GD \gavg+\gavgx{2}GDD\gavgx{2}$&  0.64 & 373.9 & 11.0  \\
\hline 
\hline
PdD, \ \ \ $T_c^{\mathrm{exp}} =$ 11 K\cite{stritzker1972superconductivity},  10 K\cite{schirber1974concentration}  \\
\hline
Self-energy & $\lambda$ & $\omega_{\mathrm{log}}$ (cm$^{-1}$) & $T_c$ (K)  \\ 
\hline
$gGDg$ & 0.48 & 294.8 & 2.5  \\
$\gavg GD \gavg$ & 0.33 & 272.4 & 0.09  \\
$gGDg+\overset{\sss{(2)}}{g}GDD\overset{\sss{(2)}}{g}$& 1.11 & 363.9 & 36.9 \\
$\gavg GD \gavg+\gavgx{2}GDD\gavgx{2}$&  0.72 & 309.8 & 13.0  \\
\hline
\hline
\end{tabular}
\caption{Calculated superconducting critical temperatures with Allen-Dynes modified equation~\cite{PhysRevB.12.905} with different electron self-energies, labeled as in Fig. \ref{fig:main}, assuming  $\mu^*=0.13$, for both PdH and PdD. The results are compared with the experimental values~\cite{stritzker1972superconductivity,schirber1974concentration}. 
The phonon propagator always includes anharmonic effects within the SSCHA.
The values obtained for the electron-phonon coupling constant $\lambda$ and average logarithmic phonon frequency, $\omega_{\mathrm{log}}=\exp\left( 2/\lambda \int_0^{\infty}d\omega \alphasqF(\omega)\log\omega/\omega \right)$, are also provided.}
\label{Tab:tc_values}
\end{table}
In this work, we focus on the first two orders ($n=1,2$), as the results shown in Fig.~\ref{fig:LambdaFD_PdH} suggest that contributions involving the third derivative ${\overset{\ssst{(3)}}{g}}$ are comparatively small, and therefore expected to be even smaller for $\avg{\overset{\ssst{(3)}}{g}}$. As shown in Fig. \ref{fig:main}, even if a different DFT functional is used here~\cite{footnote_SI}, the electron-phonon coupling constant obtained at the first order with bare vertices is consistent with previous calculations~\cite{ErreaB2013}, confirming the interpretation that non-perturbative anharmonic effects make $\lambda$ larger for PdD than for PdH, but with values too small to explain the observed $T_c$ (see Table \ref{Tab:tc_values}). As anticipated, including the averaged vertices in the electron self-energy significantly reduces $\lambda$, from 0.42 to 0.27 for PdH and from 0.48 to 0.33 for PdD. Consequently, the superconducting critical temperature decreases even further relative to the linear-order calculation, becoming nearly vanishing for both isotopes. Remarkably, the second-order electron-phonon coupling constant calculated with bare $\overset{\sss{(2)}}{g}$ vertices yields very large values for $\lambdax{2}$, surprisingly larger than the linear term, as already suggested by the finite-difference calculations presented in Fig. \ref{fig:LambdaFD_PdH}. Adding both linear and second-order terms with bare vertices yields a very large value of $\lambda$ of 1.10 for PdH and practically the same value for PdD, 1.11. Such a huge increase in the electron-phonon coupling constant yields an overestimated $T_c$ and erases any trace of an inverse isotope effect. The situation is corrected when the electron-phonon coupling constant is calculated considering first- and second-order terms with average $\gavg$ and $\gavgx{2}$ vertices, which bring the total $\lambda$ to 0.64 for PdH and 0.72 for PdD, not far from the values estimated within the semi-empirical Gaspari and Gyorffy theory~\cite{Papaconstantopoulos1978Band}. As shown in Table \ref{Tab:tc_values}, in this latter case the $T_c$ values are close to the experimental values, with a clear inversion of the isotope effect.

While the inclusion of higher-order terms and average vertices affects the average phonon frequencies (see Table \ref{Tab:tc_values} for the average logarithmic phonon frequencies), the largest impact on the superconducting critical temperatures comes from the sensitivity of $\lambda$ on these effects. The main reason for the large enhancement of the electron-phonon coupling constant when second-order terms are included is that a large contribution to $\overset{\sss{(2)}}{\alphasqF}(\omega)$ appears from the double phonon contribution of Pd-character acoustic modes, which yield a large contribution to $\lambdax{2}(\omega)$ below 400 cm$^{-1}$ (see middle panel of Fig. \ref{fig:main}). Mixed  Pd-character  acoustic and H-character optical double phonon terms do not have a large contribution to $\overset{\sss{(2)}}{\alphasqF}(\omega)$, as noted by the tiny peaks in the 400-900 cm$^{-1}$ range for PdH and 400-600 cm$^{-1}$ for PdD. Double H-character optical modes have a large contribution to $\overset{\sss{(2)}}{\alphasqF}(\omega)$ above 900 cm$^{-1}$ for PdH and above 600 cm$^{-1}$ for PdD, but, when non-perturbative average vertices are considered, their contribution to $\lambdax{2}(\omega)$ is largely suppressed. This remarks that, as expected due to their large ionic displacements, the H-character contributions to $\overset{\sss{(1)}}{\alphasqF}(\omega)$ and $\overset{\sss{(2)}}{\alphasqF}(\omega)$ are the ones that are mostly affected by the inclusion of average vertices. Remarkably, the total $\alphasqF(\omega)$ exhibits substantial spectral weight at frequencies exceeding the maximum one-phonon frequency. This high-energy contribution originates from second-order processes. Given that the Eliashberg function determines the electronic transport as well as the optical conductivity of a material, the large non-linear contribution could be inferred directly from tunneling~\cite{McMillan1965Lead,Schackert2015Local,Dolgov2003Extraction} or infrared reflectivity~\cite{MARSIGLIO1998Inversion,capitani2017spectroscopic,Borinaga_2018} measurements, opening novel avenues in the interpretation of experimental data.    


In conclusion, non-linear electron-phonon effects strongly enhance the superconducting critical temperature in palladium hydrides. A perturbative treatment including bare second-order terms breaks down, overestimating $T_c$ and suppressing the isotope anomaly. The inverse isotope effect and quantitative agreement with experiments are recovered only when a non-perturbative treatment of the electron-phonon interaction—where the vertices include an infinite resummation accounting for quantum and thermal ionic fluctuations—is combined with a fully anharmonic quantum phonon theory such as the SSCHA. These findings suggest that strong non-linear electron-phonon effects may be widespread in superconducting hydrides~\cite{novoa2025realspaceunderstandingelectronphononcoupling} and could play a central role in slightly doped ferroelectric superconductors, where anharmonicity is essential~\cite{Edge2015Quantum,Gastiasoro2023Generalized}.

\vspace{0.5cm}

We acknowledge fruitful discussions with Philip Allen,
Lorenzo Paulatto, and Matteo Calandra. We received financial support from the European Research Council (ERC) under the European
Union’s Horizon 2020 research and innovation program
(Grant Agreement No. 802533); 
the PID2022-142861NA-I00  project funded by MICIU/AEI/10.13039/501100011033 and FEDER, UE;
the Department
of Education, Universities and Research of the Eusko
Jaurlaritza and the University of the Basque Country UPV/EHU (Grant No. IT1527-22); and Simons Foundation through the Collaboration on New Frontiers in Superconductivity (Grant No. SFI-MPS-NFS-00006741-10). Computational resources were granted by the Red Española de Supercomputación (Grant No. FI-2022-3-0010).



\clearpage
\onecolumngrid

\setcounter{section}{0}
\setcounter{equation}{0}
\setcounter{figure}{0}
\setcounter{table}{0}

\renewcommand{\thesection}{S\arabic{section}}
\renewcommand{\theequation}{S\arabic{equation}}
\renewcommand{\thefigure}{S\arabic{figure}}
\renewcommand{\thetable}{S\arabic{table}}

\section*{Supplemental Material}

\section*{I. Computational Details}
We performed density-functional theory (DFT) calculations using the QUANTUM-ESPRESSO package~\cite{Giannozzi2009}, and we employed optimized norm-conserving pseudopotentials (ONCCPSP)~\cite{PhysRevB.88.085117,VANSETTEN201839}, together with the Perdew–Burke–Ernzerhof (PBE) parametrization of the generalized gradient approximation (GGA) for the exchange–correlation functional~\cite{PhysRevLett.77.3865}.  Self-consistent calculations were performed on a $2\times 2 \times 2$ supercell. A plane-wave cutoff of $65$~Ry was used, together with a $12 \times 12 \times 12$ k-point mesh for the Brillouin-zone integrals required to compute the self-consistent Kohn–Sham potentials. For the unit cell of PdH and PdD we used the lattice parameters  $a_{\text{PdH}}=7.735\,\text{a.u.}$ and $a_{\text{PdD}}=7.725\,\text{a.u.}$, respectively,  as estimated in Ref.~\citenum{ErreaB2013} by relaxing the structure while  accounting for anharmonic and quantum effects in the nuclear dynamics, and in good agreement with experimental measurements.

\begin{figure}[t!]
\includegraphics[width=\linewidth]{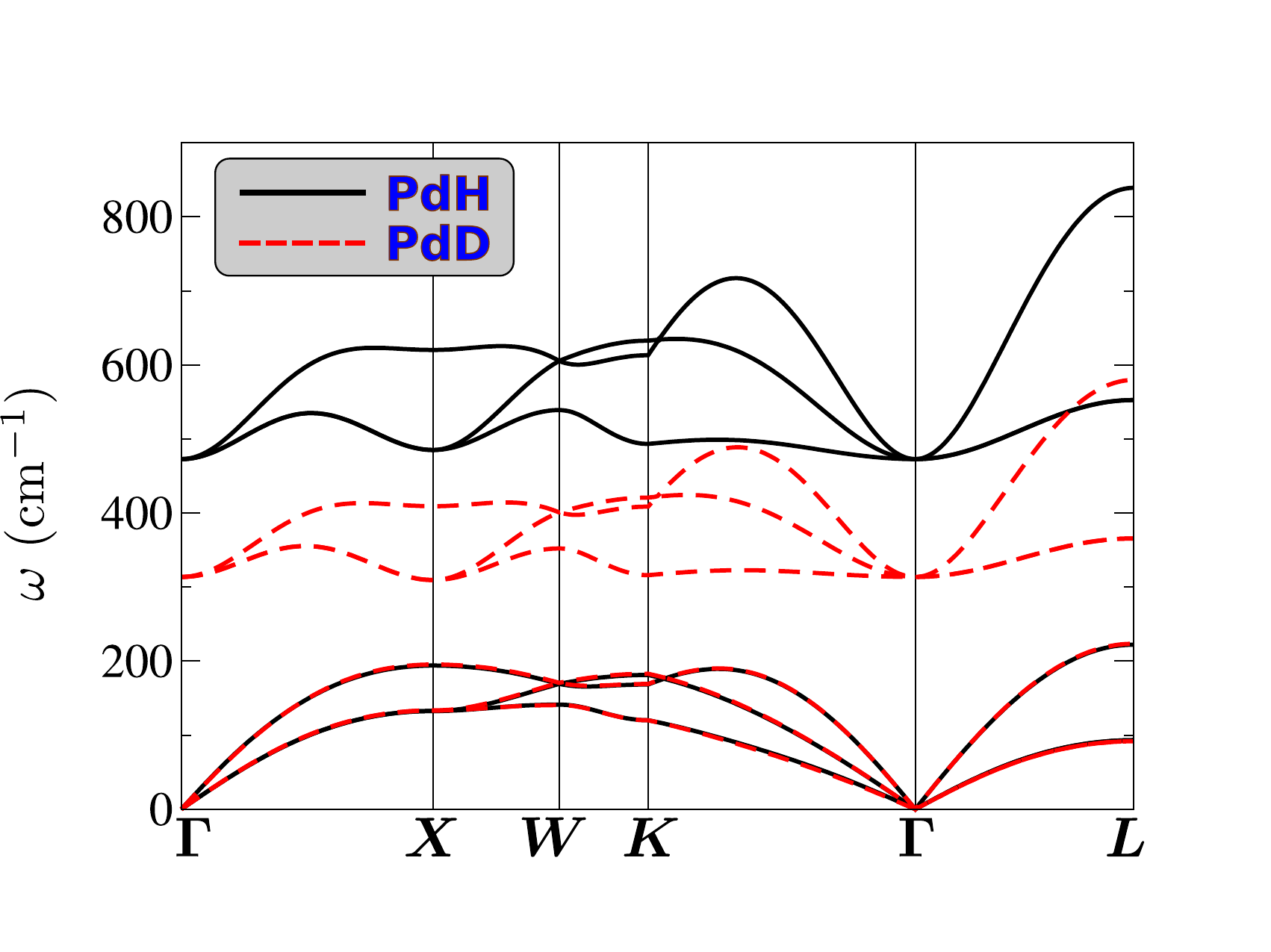}
\caption{Self-consistent harmonic phonon dispersions of PdH and PdD.}
\label{fig:SCHA_disp}
\end{figure}

The effective harmonic description of the nuclear dynamics, including quantum and anharmonic effects, was obtained using the Stochastic Self-Consistent Harmonic Approximation (SSCHA)~\cite{Monacelli_2021} in a $2\times 2\times 2$ supercell. The resulting self-consistent harmonic phonon dispersion along a high-symmetry path is shown in Fig.~\ref{fig:SCHA_disp}. 

\section*{II. Convergence Study}
The first- and second-order electron–phonon coupling constants, 
$\lambdax{1}$ and $\lambdax{2}$, were first computed from the 
$gGDg$ and $\overset{\sss{(2)}}{g}GDD\gx{2}$ self-energies, respectively, through equations
\begin{align}
\overset{\sss{(1)}}{\alphasqF}(\omega)=
&\Nc\Nf\,\sum_{\mu}
\AAvg{\,\left|\overset{\sss{(1)}}{g}{}^{n m\bk}_{\mu}\right|^2\,}_{\substack{{n\bk}\\{m\bk}}\in\text{FS}}\delta\left(\omega-\omega_{\mu}\right)
&&\overset{\sss{(1)}}{\lambda}=2\int_0^{+\infty}\mkern-10mu d\omega\,\frac{\overset{\sss{(1)}}{\alphasqF}(\omega)}{\omega}\\
\overset{\sss{(2)}}{\alphasqF}(\omega)=
&2\Nc\Nf\,\sum_{\mu\nu}
\AAvg{\,\left|\overset{\sss{(2)}}{g}{}^{n m\bk}_{\mu\nu}\right|^2\,}_{\substack{{n\bk}\\{m\bk}}\in\text{FS}}\delta\left(\omega-\omega_{\mu}-\omega_{\nu}\right)
&&\overset{\sss{(2)}}{\lambda}=2\int_0^{+\infty}\mkern-10mu d\omega\,\frac{\overset{\sss{(2)}}.{\alphasqF}(\omega)}{\omega}
\end{align}
The derivatives of the self-consistent KS potential entering in these equations were calculated by finite difference.

\begin{figure}[t!]
\includegraphics[width=1.0\linewidth]{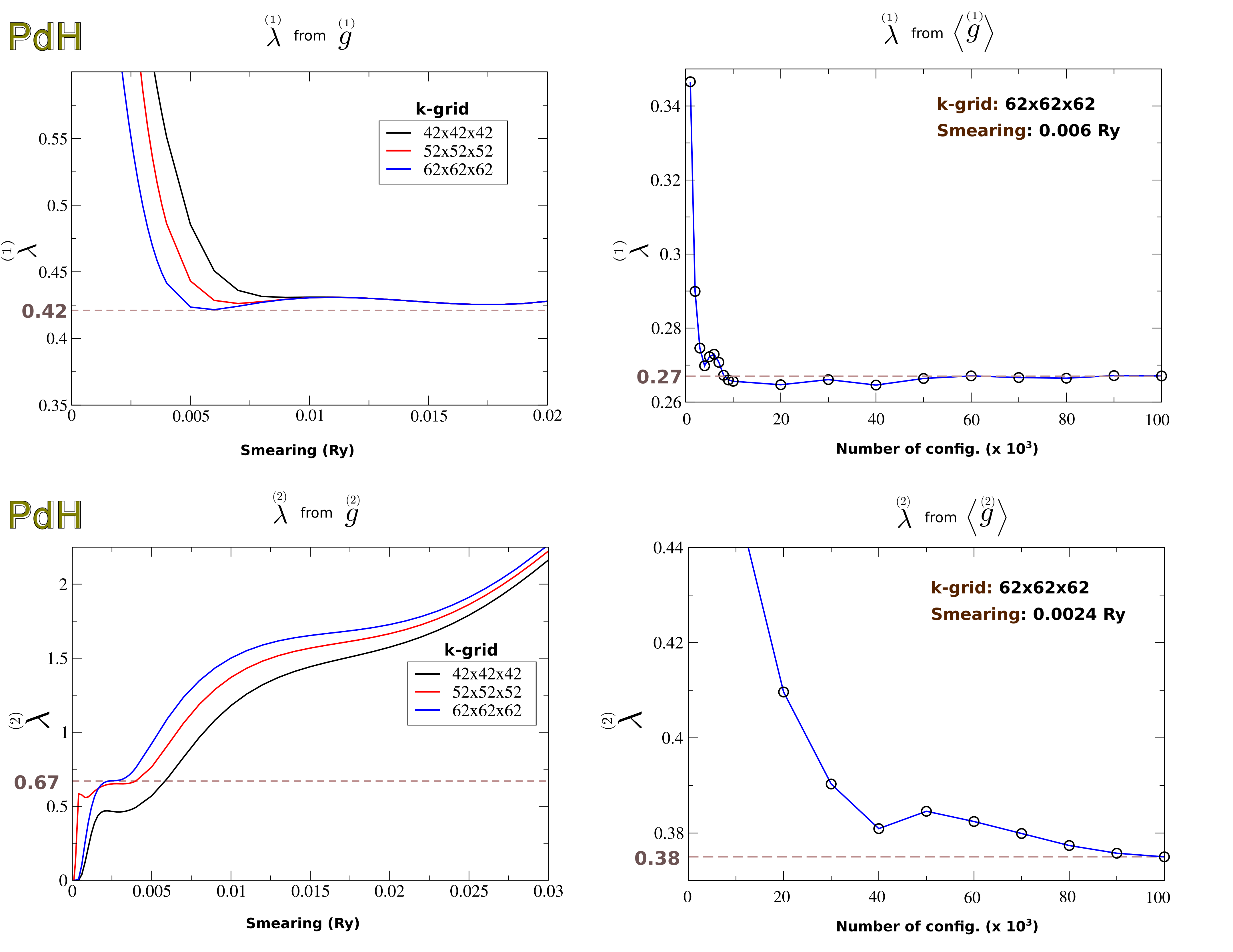}
\caption{Left panels: values of $\overset{\sss{(1)}}{\lambda}$ and 
$\overset{\sss{(2)}}{\lambda}$ for PdH, computed from the $gGDg$ and 
$\overset{\sss{(2)}}{g}GDD\gx{2}$ self-energies, respectively, 
as a function of the $k$-point grid used for the Brillouin-zone integration 
and of the Gaussian smearing employed to approximate the delta functions.
Right panels: values of $\overset{\sss{(1)}}{\lambda}$ and 
$\overset{\sss{(2)}}{\lambda}$ for PdH, computed from the 
$\gavg GD\gavg$ and $\gavgx{2}GDD\gavgx{2}$ self-energies, respectively, 
using the selected $k$-point grid and Gaussian smearing, 
as a function of the stochastic population size employed to evaluate the averaged vertices.
}
\label{fig:PdH_Lambda}
\end{figure}

\begin{figure}[t!]
\includegraphics[width=1.0\linewidth]{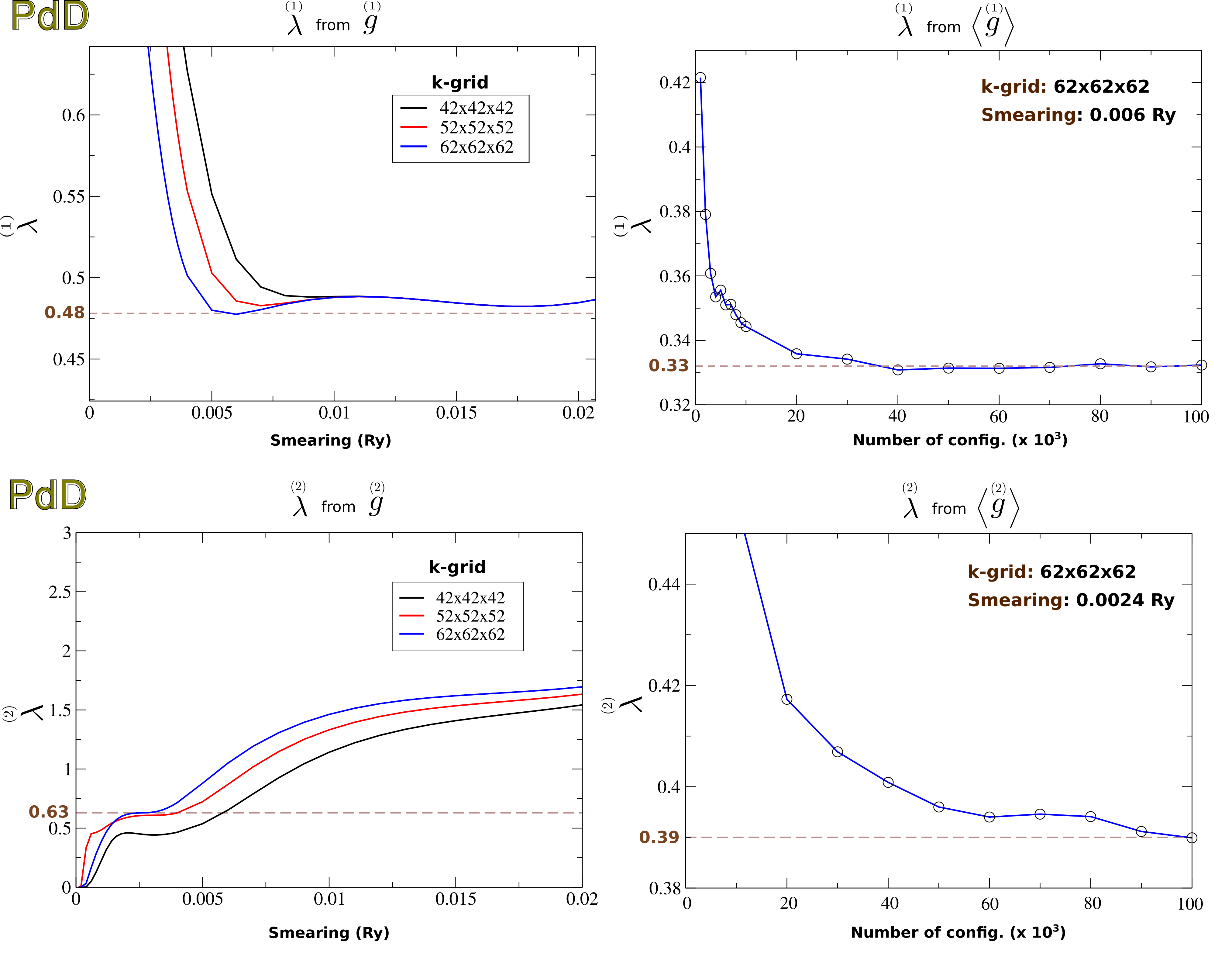}
\caption{Left panels: values of $\overset{\sss{(1)}}{\lambda}$ and 
$\overset{\sss{(2)}}{\lambda}$ for PdD, computed from the $gGDg$ and 
$\overset{\sss{(2)}}{g}GDD\gx{2}$ self-energies, respectively, 
as a function of the $k$-point grid used for the Brillouin-zone integration 
and of the Gaussian smearing employed to approximate the delta functions.
Right panels: values of $\overset{\sss{(1)}}{\lambda}$ and 
$\overset{\sss{(2)}}{\lambda}$ for PdD, computed from the 
$\gavg GD\gavg$ and $\gavgx{2}GDD\gavgx{2}$ self-energies, respectively, 
using the selected $k$-point grid and Gaussian smearing, 
as a function of the stochastic population size employed to evaluate the averaged vertices.}
\label{fig:PdD_Lambda}
\end{figure}

The supercell $k$-point grid used to perform the Brillouin-zone integration entering the double Fermi-surface averages, together with the Gaussian smearing employed to approximate the delta functions, were determined through a systematic convergence analysis. The results of this analysis are shown in the left panels of 
Figs.~\ref{fig:PdH_Lambda} and~\ref{fig:PdD_Lambda} for PdH and PdD, respectively. While the first-order value of $\lambda$ is converged with the number of  $k$-points, the convergence of $\lambdax{2}$ is more challenging, and the parameters used in our calculations suggest that it is converged up to $\sim 0.15$.

The same $k$-point grid and smearing parameters were then adopted to compute 
the electron–phonon coupling constants from the 
$\gavg GD\gavg$ and $\gavgx{2}GDD\gavgx{2}$ self-energies following the stochastic method outlined in the paper  submitted in parallel to this
Letter. 
The convergence with respect to the stochastic population size used 
to evaluate the averaged vertices is reported in the 
right panels of Figs.~\ref{fig:PdH_Lambda} and~\ref{fig:PdD_Lambda} for PdH and PdD, respectively.

\section*{III. Single- and two-phonon spectral quantities}
\begin{figure}[t!]
\includegraphics[width=0.7\linewidth]{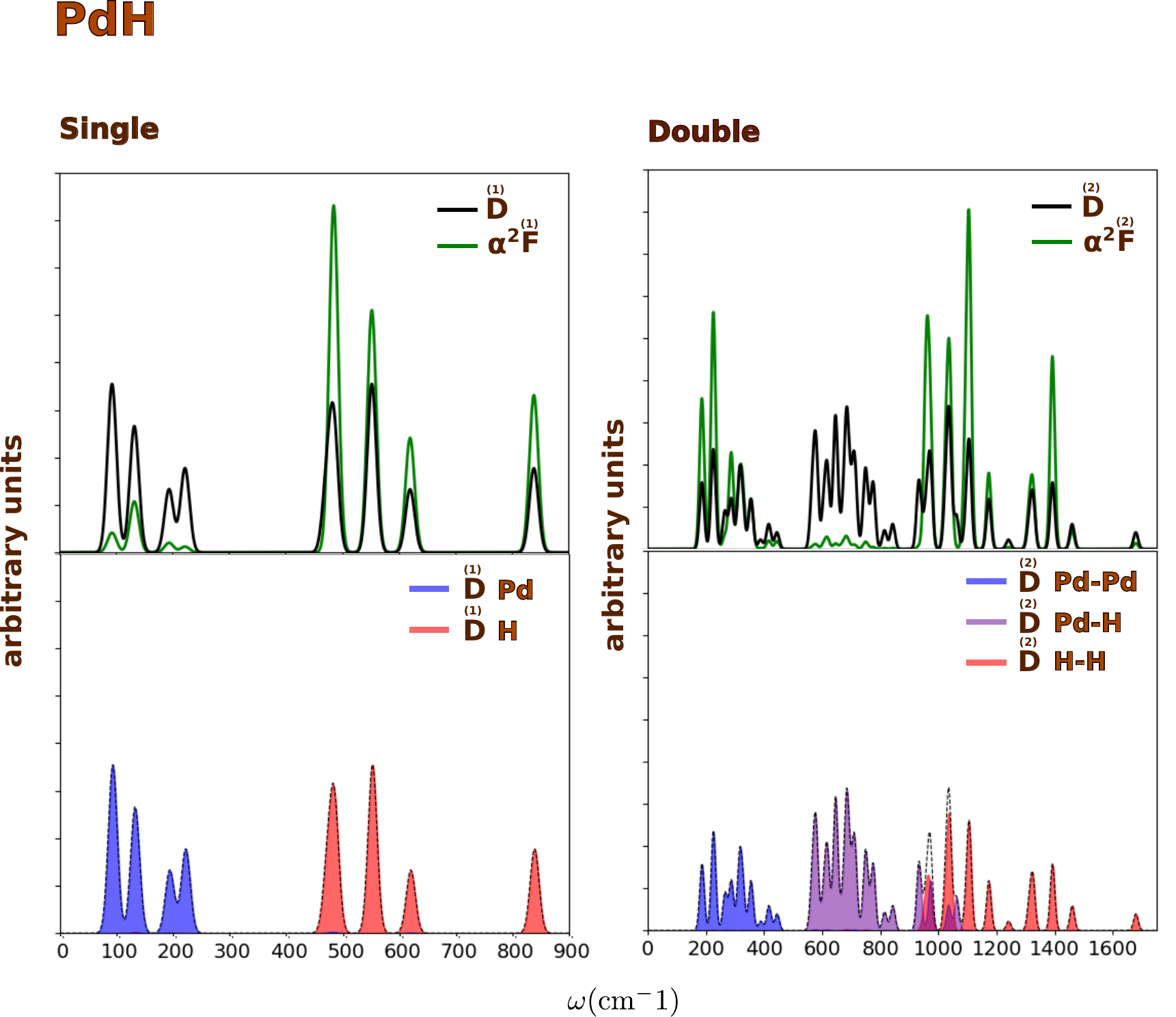}
\caption{Single- and two-phonon spectral quantities for PdH. 
Left panels: top, single-phonon density of states ($\overset{\ssst{(1)}}{D}$) 
and first-order Eliashberg spectral function ($\alpha^2\overset{\ssst{(1)}}{F}$); 
bottom, atom-projected single-phonon density of states. 
Right panels: top, two-phonon density of states ($\overset{\ssst{(2)}}{D}$) 
and second-order Eliashberg spectral function ($\alpha^2\overset{\ssst{(2)}}{F}$); 
bottom, atom-projected two-phonon density of states. 
For clarity, the density of states and Eliashberg spectral functions 
have been normalized to the same total integrated value.}
\label{fig:SpecPdH}
\end{figure}

\begin{figure}[t!]
\includegraphics[width=0.7\linewidth]{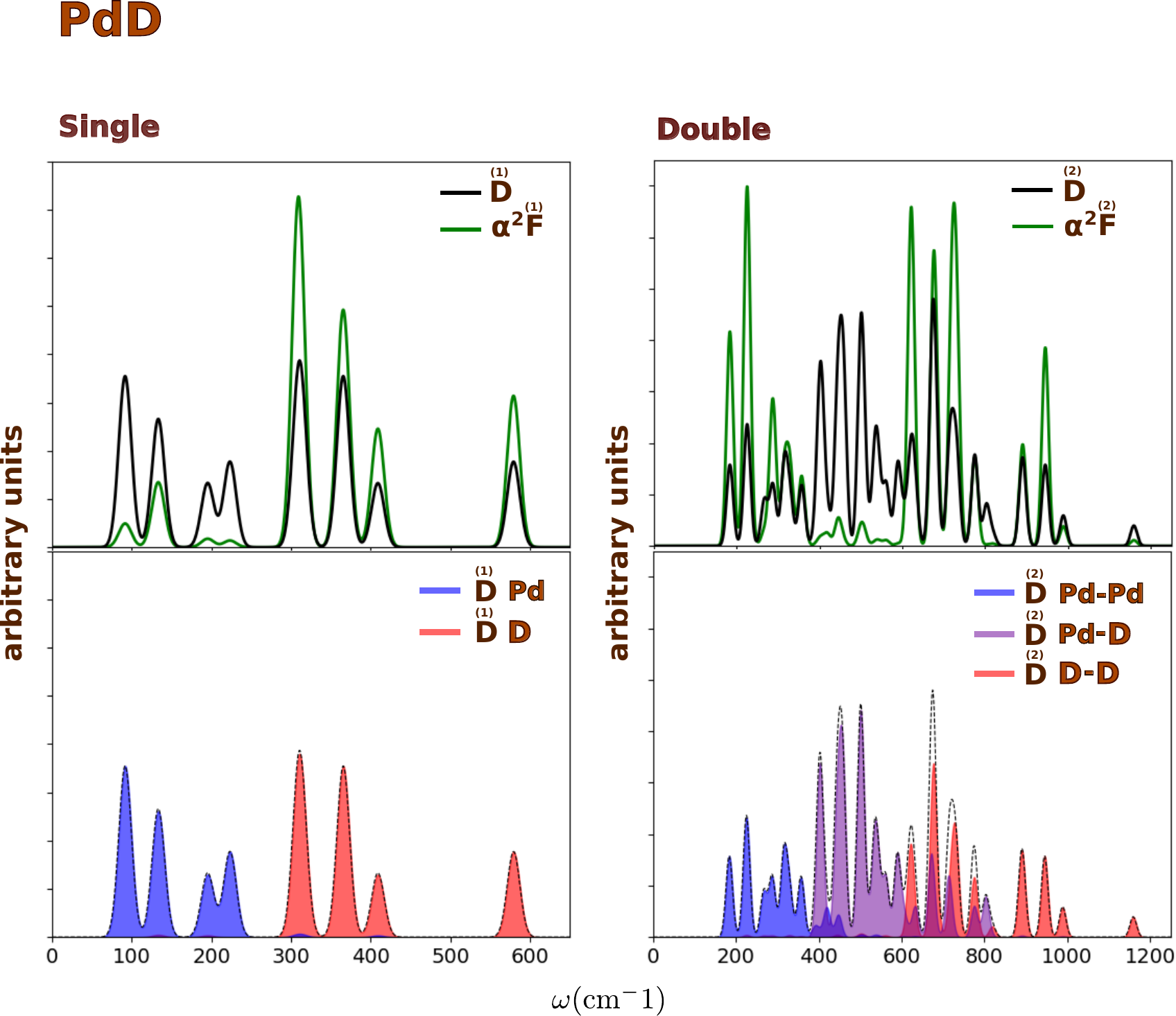}
\caption{Single- and two-phonon spectral quantities for PdD. 
Left panels: top, single-phonon density of states ($\overset{\ssst{(1)}}{D}$) 
and first-order Eliashberg spectral function ($\alpha^2\overset{\ssst{(1)}}{F}$); 
bottom, atom-projected single-phonon density of states. 
Right panels: top, two-phonon density of states ($\overset{\ssst{(2)}}{D}$) 
and second-order Eliashberg spectral function ($\alpha^2\overset{\ssst{(2)}}{F}$); 
bottom, atom-projected two-phonon density of states. 
For clarity, the density of states and Eliashberg spectral functions 
have been normalized to the same total integrated value.}
\label{fig:SpecPdD}
\end{figure}

Figures~\ref{fig:SpecPdH} and~~\ref{fig:SpecPdD} show, for PdH and PdD, the single- and double- phonon density of states
\begin{align}
&\overset{\ssst{(1)}}{D}(\omega)=\sum_\mu\delta(\omega-\omega_\mu)\\
&\overset{\ssst{(2)}}{D}(\omega)=\sum_{\mu\nu}\delta(\omega-\omega_\mu-\omega_\nu)\,, 
\end{align}
compared  with the first and second order Eliashberg spectral functions $\overset{\sss{(1)}}{\alphasqF}(\omega)$ and $\overset{\sss{(2)}}{\alphasqF}(\omega)$.
The atom-projected density of states is also shown, and clearly separates Pd-dominated acoustic modes  from hydrogen-derived optical modes at higher frequencies.  The Eliashberg spectral functions do not simply follow  the phonon density of states. A redistribution of spectral weight is observed, indicating that the superconducting pairing interaction  is not governed solely by the phonon density of states, but is strongly  influenced by the coupling strength of specific anharmonic modes.




\FloatBarrier
\clearpage
\flushbottom

\end{document}